# Full-span reversible space-time birefringence

*Chenhui Yu[1,2], Guanyi Zhu[1,2], Mingliang Xu[1,3], Fei He[1,2]\*, Liwei Song[1,2], Ye Tian[1,2], Yuxin Leng[1,2], Ruxin Li[1,2,4]*

C. Yu, G. Zhu, M. Xu, F. He, L. Song, Y. Tian, Y. Leng, R. Li
[1]State Key Laboratory of Ultra-intense Laser Science and Technology, Shanghai Institute of Optics and Fine Mechanics, Chinese Academy of Sciences, Shanghai 201800, China

[2]Center of Materials Science and Optoelectronics Engineering, University of Chinese Academy of Sciences, Beijing 100049, China
E-mail: hefei@siom.ac.cn (F.H.)

M. Xu
[3]School of Physical Sciences, University of Science and Technology of China, Hefei 230026, China

R. Li
[4]Zhangjiang Laboratory, Shanghai 201210, China

Funding: This work was funded by the National Natural Science Foundation of China No.12388102 (RL, YL), the CAS Pioneer Hundred Talents Program (FH), the Zhangjiang Laboratory Youth Innovation Project ZJYI2022A01 (FH), S202420005 (FH), and the Shanghai Science and Technology Committee Program No. 23560750200 (YL).

**Abstract**: Birefringence, the polarization-dependent splitting of light in anisotropic crystals, enables diverse optical phenomena and advanced functionalities such as optical communication, nonlinear optics, and quantum optics. However, conventional methods for controlling birefringence typically rely on engineering the optical crystal structure or applying external stimuli such as electric fields, mechanical stress or thermal variations, which are often constrained by limited tunability, challenges in integration with compact photonic devices or slow response time. Here, we introduce a new degree of freedom to manipulate the birefringence of light propagation in optical crystals through programming the spatiotemporal spectral phase of the incident light wave. We demonstrate this approach achieves continuous tuning of birefringence across a spectrum more than 100 times broader than that achievable with conventional birefringence tuning, spanning from positive through zero to negative values, irrespective of the crystal's optical sign and without inherent physical limitations. This unique optical behavior provides a versatile platform for investigating the complex dynamics of wave flow in anisotropic media, while the broad tunability of this space-time birefringence will spur innovations in ultrafast optical manipulation, optical computation, and quantum information processing− applications that demand rapid and flexible device reconfiguration.

## 1. Introduction

Birefringence is an optical property observed in anisotropic materials, wherein light travels at different velocities along distinct crystallographic axes. As a result, a single incident light beam separates into two orthogonally polarized eigenwaves, *e.g.*, the ordinary (o) wave and the extraordinary (e) wave in uniaxial crystals.[1, 2] In such materials, the refractive index ($n$) varies with both the polarization and propagation direction of light, resulting in different velocities and phase retardation for the orthogonally polarized components. Conventionally, birefringence has been controlled via external perturbations including electric fields (Pockels effect), magnetic fields (Faraday effect), mechanical stress, or temperature changes.[3] This property enables precise modulation of light polarization and phase, rendering birefringence highly valuable in scientific research and industrial applications. In optical communications, birefringent materials are employed in key components such as waveplates, polarizers, isolators/circulators, and dispersion compensators.[4] In material characterization, interferometry and ellipsometry exploit birefringence to measure strain and optical anisotropy.[5-7] Its tunable and polarization-sensitive nature continues to propel advancements in emerging fields such as optical computing and quantum optics. For instance, birefringence serves as a key mechanism for implementing optical logic gates in all-optical computing systems,[8, 9] and enables precise manipulation of photonic qubits for quantum state encoding in quantum information processing.[10-13]

Recently, non-diffractive space-time (ST) light represent specialized optical wave packets that preserve their spatial and temporal profiles over extended distances, defying classical diffraction and dispersion.[14-24] These one- or two-dimensional structures arise from spatiotemporal coupling, which entangles spatial and temporal degrees of freedom in the wavefunctions. Unlike conventional light, ST light has a non-separable spectral structure−its spatial and temporal components are inherently linked. This non-separability creates a unique spatiotemporal profile with non-differentiable angular dispersion, where each spatial frequency corresponds to a single prescribed wavelength. Mathematically, this correlation restricts the spectral support to the intersection of the light-cone and a spectral plane defined by $\Omega = (k_z - k_o)c\tan\theta$, where $\Omega = \omega - \omega_o$ is the angular frequency detuning from a carrier frequency $\omega_o$, $k_o = \omega_o/c$ is the wavenumber, and $\theta$ is the spectral tilt angle. This configuration enables propagation invariance: the envelope preserves its shape as $\psi(x, z; t) = \psi(x, 0; t - z/v_g)$, propagating at a tunable group velocity $v_g$ that can be faster or slower than the light speed in vacuum ($c$) without violating relativistic causality. Compared to Gaussian beams, ST light exhibits tunable group velocities,[16] propagation-invariant,[17] dispersion-free propagation in dispersive media,[24] the 'flying focus' effect,[25] and exotic refraction beyond Snell's Law and Fermat's Principle.[18] These properties can be harnessed to program electromagnetic wave dynamics in both space and time, enable spatiotemporal engineering of material properties and particle behaviors,[23, 26] and provide a platform for simulating quantum and relativistic phenomena under well-controlled laboratory conditions.[20, 27, 28]

Although research on ST light in isotropic media has made significant progress, its behavior in anisotropic media—particularly in birefringent optical crystals—remains poorly understood. This is due to the complex interplay between the entangled spatiotemporal properties of ST light and the direction-dependent dielectric response of multi-axis crystals. Anisotropic materials have refractive indices that vary with propagation direction and polarization, causing effects like birefringence and polarization-dependent dispersion.[2, 3] These properties complicate predictions of how ST light propagates, evolves, and maintains coherence in such media. This gap is critical because anisotropy couples polarization and propagation direction, disrupting the spectral correlations that define ST light and leading to effects like ST birefringence. In this case, different polarization components experience different dispersion or diffraction behaviors, causing the pulse splitting or deform during propagation. Understanding these effects is essential for modeling ST light in realistic settings, particularly in precision optical systems and materials with intrinsic anisotropy, such as crystals, fibers or metamaterials.[1, 23, 29, 30]

In this study, we advance the understanding of ST light propagation in anisotropic media by elucidating its birefringence characteristics in optical crystals. We propose a 'double light-cone' model to describe the birefringent spectrum of ST light in uniaxial crystals. Unlike conventional birefringence, which rely solely on the (bi-)refractive index of the medium, ST birefringence can be actively tuned by adjusting the incident angle and the spectral phase of the incident ST light. Strikingly, in positive uniaxial crystals, the group velocity of o-ST light can exceed, fall below, or equal that of the e-ST component—reversals of this behavior are observed in negative uniaxial crystals. We further experimentally confirmed these predictions through interferometric group-delay measurements in a YVO$_4$ crystal. By simply adjusting the spectral phase pattern on a spatial light modulator (SLM), ST birefringence can be programmed broadly, providing a new degree of freedom for real-time control of light manipulation and light-matter interactions without altering the crystal's physical properties. Our findings deepen the understanding of wave propagation in anisotropic media, and hold promise for applications in reconfigurable, compact photonic devices and optical systems.

## 2. Results

### 2.1. 'Double light-cone' representation of light propagation in uniaxial crystals

For ST light in the free space, a spatiotemporal light-cone representation has been proposed to visualize its spectral properties based on the dispersion relation $k_x{}^2 + k_z{}^2 = n^2(\omega/c)^2$, where $n$ is the refractive index of an isotropic, nondispersive medium, $k_x$ and $k_z$ are the transverse and axial wavevector components along $x$ and $z$, respectively, $\omega$ is the angular frequency, $c$ is the speed of light in vacuum. For simplicity, the field along $y$ is assumed to be uniform. This model works well in describing the spatiotemporal spectrum of physically realizable optical fields under the narrow-band paraxial (NBPA) approximation in isotropic media.[31] However, in uniaxial crystals, the o-light follows Snell's law, while the e-light deviates due to a direction-dependent refractive index relative to the optical axis (OA). We propose a 'double light-cone' representation to describe both types of light propagation in anisotropic media. For clarity, we consider a uniaxial crystal with an arbitrary OA direction. The propagation dynamics of o-light follows the isotropic dispersion relation. In contrast, the refractive index of e-light varies explicitly with the angle $\beta$ between the wave vector and the OA, as given by $1/n_e{}^2(\beta) = \cos^2\beta/n_o{}^2 + \sin^2\beta/n_e{}^2$, where $n_o$ and $n_e$ are the ordinary and extraordinary principal refractive indices, respectively. Thus, the dispersion relation for e-light is expressed as $k_x{}^2 + k_z{}^2 = n_e{}^2(\beta)(\omega/c)^2$, for simplicity, the dispersion of the refractive index was neglected.

**Figure 1A** shows the refractive index ellipsoid of a uniaxial crystal with the orientation of the OA is specified by two angles: $\varphi$, the tilt angle of the OA relative to the $y$-axis, and $\alpha$, the azimuthal orientation of the OA projection onto the $(x, z)$-plane relative to the $z$-axis. Together, these angles determine the wavefront symmetry and medium anisotropy, influencing both the group velocity and the spatiotemporal coupling of ST light. Using this geometry, for describing light propagating along $z$ direction, the light-cone dispersion relation for e-light can be formulated by transforming the standard dispersion relation into a coordinate system aligned with the OA as

$$k_x^2 + k_z^2 = \frac{n_o^2 n_e^2}{n_o^2 \sin^2\beta + n_e^2 \cos^2\beta}\left(\frac{\omega}{c}\right)^2, \tag{1}$$

where $\beta = \cos^{-1}(\frac{k_x\sin\varphi\sin\alpha + k_z\sin\varphi\cos\alpha}{\sqrt{k_x^2 + k_z^2}})$.

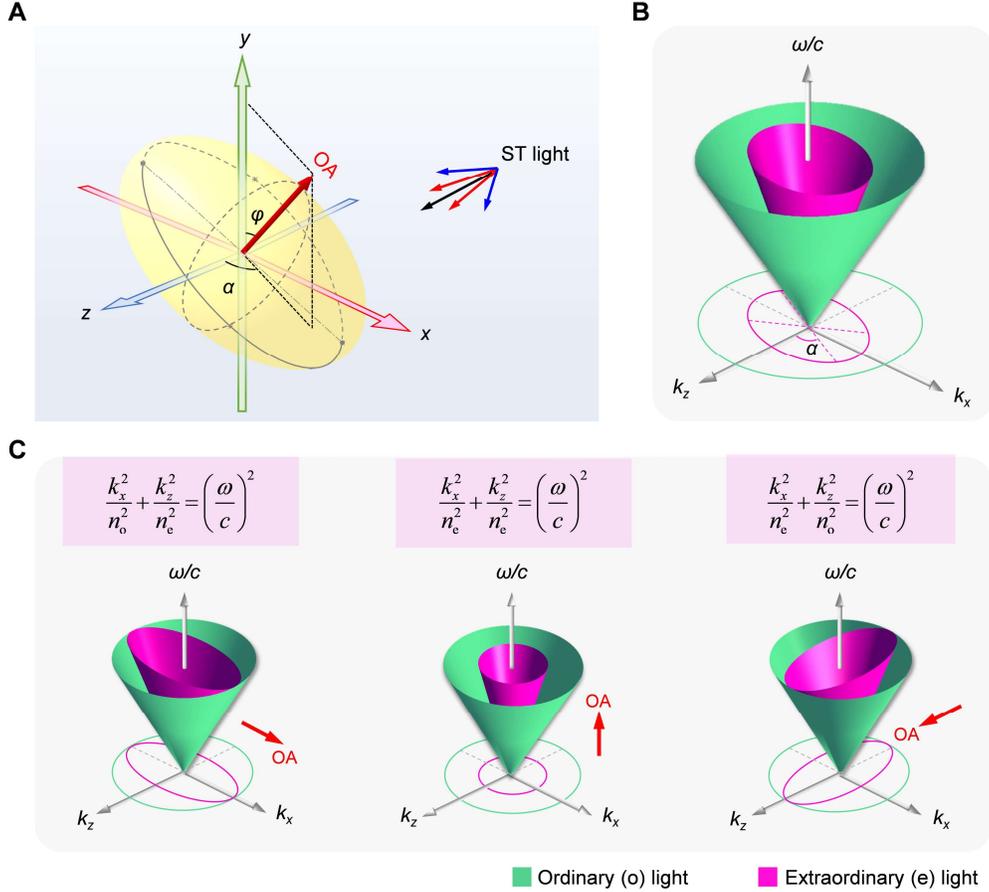

**Figure 1. 'Double light-cone' representing the light propagation in a uniaxial optical crystal.** (**A**) Refractive index ellipsoid of a negative uniaxial crystal, with light propagation along $z$-direction. The red solid arrow indicates the direction of the OA. $\alpha$ denotes the angle between the projection of the OA in the $(x, z)$-plane and the $z$-axis, $\varphi$ is the angle between the OA and the $y$-axis. (**B**) Double light-cone surfaces representation of the birefringent dispersion of light propagating in a uniaxial negative crystal as shown in (A). The green and pink conical surfaces represent the dispersion relation of o- and e-light in the spatiotemporal spectral space, respectively. (**C**) Double light-cone surfaces representation of the birefringent dispersion of light propagating in a uniaxial negative crystal with OA typically orientated to $x$, $y$ and $z$ directions, respectively. The red arrows indicate

the directions of the OA. The dispersion relation equations of e-light corresponding to pink cones with each OA orientation are listed accordingly.

The dispersion relations for o- and e-light together form a general 'double light-cone' representation for light propagating in a uniaxial negative crystal, as shown in **Figure 1B**. The green and pink cones represent the dispersion relations of the o- and e-light, respectively, with the dispersion of e-light following **Eq. (1)**. When projected onto the $(k_x, k_z)$-plane, the pink cone forms an elliptical contour rotated by an angle $\alpha$ relative to the coordinate axes, reflecting the optical anisotropy of the crystal. Notably, the major and minor axes of the pink cone are proportional to the $n_e(90°-\varphi)$ and $n_e$, respectively. **Figure 1C** displays typical 'double light-cone' configurations when the OA is aligned with the $x$-, $y$-, and $z$-directions. In each case, the polarization modes degeneracy and splitting arise from the crystal's anisotropy. These light-cones provide a geometric framework for understanding wave propagation in uniaxial crystals, linking mathematical expressions to intuitive visualizations in momentum-frequency space.

## 2.2. ST light deviates from the classical birefringent law in crystals

For normally incident ST light into isotropic media, both energy conservation and lateral momentum conservation ensures the preservation of the spectral trajectory in the $(k_x, \omega/c)$-plane.[18, 32, 33] Under the NBPA approximation, this trajectory takes the form of a parabola, with curvature as the invariant, expressed as $n(n-n_g)$, where $n_g = \cot\theta$ is the group index and $\theta$ is the spectral tilt angle. The group velocity is given as $v_g = c\tan\theta$, or equivalently $v_g = c/n_g$. The invariance of the quantity $n(n-n_g)$ guarantees that this product remains constant across interfaces, despite changes in the local refractive index, enabling predictable manipulation of ST light in diverse optical systems.

In free space, the spectral projection of ST light onto the $(k_z, \omega/c)$-plane forms an oblique line at an angle $\theta$, namely spectral tilt angle, with respect to the $k_z$-axis. The spectral projection onto the $(k_x, \omega/c)$-plane remains invariant across the interface. Owing to the distinct light-cones of the o- and e-ST light, their spatiotemporal spectra differ, resulting in different projection angles $\theta_o$ and $\theta_e$ on the $(k_z, \omega/c)$-plane, respectively (**Figure 2A**). We can formulate the new birefringence invariant of ST light in uniaxial crystals similarly by separately defining the group indices, $n_{og}$ and $n_{eg}$ for o- and e-ST light (see **Text S1** and **Figure S1** in the Supporting Information). For the uniaxial crystal, the refraction invariant holds for o-ST light, following the relation $n(n-n_g) = n_o(n_o-n_{og})$, whereas for e-ST light, this invariant requires modification to account for anisotropic dispersion effects. Thus, we have the birefringent law of ST light is expressed as

$$n(n - n_g) = n_o(n_o - n_{og}) = \begin{cases} \frac{n_o^2}{n_e}(n_e - n_{eg}), & \text{OA} \mathbin{/\!/} \vec{x} \\ n_e(n_e - n_{eg}), & \text{OA} \mathbin{/\!/} \vec{y} \\ \frac{n_e^2}{n_o}(n_o - n_{eg}), & \text{OA} \mathbin{/\!/} \vec{z} \end{cases}. \tag{2}$$

**Eq. (2)** establishes a birefringent invariant for ST light undergoing normal transmission across the interface between isotropic and anisotropic media. When the OA is aligned along the $x$- or $z$-direction, the double-light-cone structure comprises a circular cone for o-light and an elliptical cone for e-light (**Figure 1C**, left and right), and the propagation invariance of both o- and e-ST components is described by **Eq. (2)**. When the OA oriented along the $y$-direction, the light propagates within the $(x, z)$-plane while remaining orthogonal to the OA, resulting in a constant refractive index equal to $n_e$ for each e-ST component. In this configuration, the birefringent dispersion manifests as a concentric double-cone structure (**Figure 1C**, middle).

The birefringent invariant relations reveal different spectral characteristics of o- and e-ST light in anisotropic media, with the corresponding dispersion differences influencing the spectral bandwidth and temporal coherence of each polarization component. By adjusting the spectral tilt angle $\theta$ of the input ST light, the effective dispersion experienced by o- and e-ST light upon entering the medium can be controlled. Since $\theta$ determines the distribution of pulse energy across wavevectors, it enables programming the group velocities, $v_{og}$ and $v_{eg}$, for o- and e-ST light, respectively, facilitating precise modulation of the relative delay or synchronization between orthogonally polarized components.

**Figure 2A-C** show the control of the group velocities of o- and e-ST light under normal incidence on a negative uniaxial crystal with its OA aligned parallel to the $z$-axis. Based on **Eq. (2)**, varying the spectral tilt angle $\theta$ of the input ST light enables modulation of the spectra of o- and e-ST light and their projections onto the $(k_z, \omega/c)$- and $(k_x, \omega/c)$-planes. On the $(k_x, \omega/c)$-plane, $\theta_o$ and $\theta_e$ are projected as spectral tilt angles for o- and e-ST light, respectively. Thus, the group index and group velocity for o-ST light is $n_{og} = \cot\theta_o$ and $v_{og} = c/n_{og} = c\tan\theta_o$, while for e-ST light, $n_{eg} = \cot\theta_e$, $v_{eg} = c/n_{eg} = c\tan\theta_e$. This allows tuning the group velocity of o-ST

light to be slower than, equal to, or faster than that of e-ST light—opposite to conventional birefringence, where the group velocities of orthogonally polarized wave components are solely determined by $n_o$ and $n_e$.

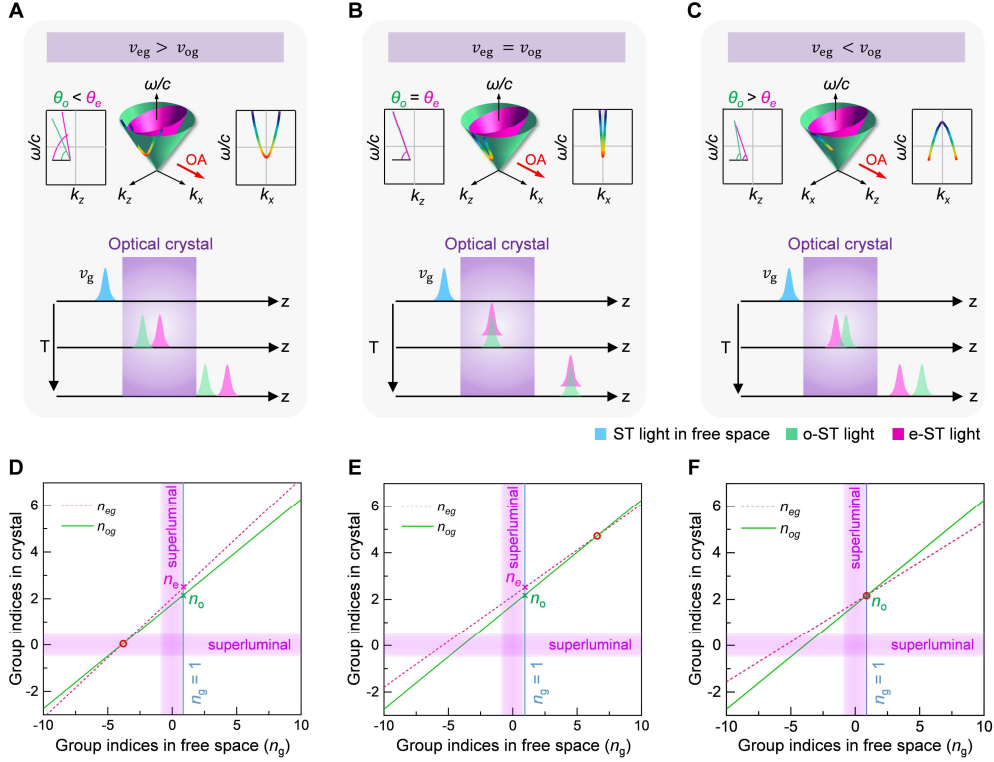

**Figure 2. Propagation dynamics of ST light in a uniaxial crystal.** (**A-C**) Concept for controlling the group velocities of o-ST and e-ST light travelling in a negative uniaxial crystal. By adjusting the spectral tilt angle of the incident ST light (blue), the propagation speed of the o-ST light (green) can be made (A) slower than, (B) equal to, or (C) faster than that of the e-ST light (pink). (**D-F**) Relationship between the free-space group index $n_g$ and the group index of o-ST ($n_{og}$, green) and e-ST ($n_{eg}$, pink) light, with the OA oriented along the $x$- (D), $y$- (E) and $z$-axis (F), respectively. The data are plotted according to **Eq. (2)**. The red circle on each plot indicates the point at which the birefringence of the ST light is canceled at a specific incident group index (*i.e.*, spectral tilt angle), marking the transition of the ST light between positive and negative birefringence, or vice versa. The solid blue lines in each plot represent incident ST light with a group index of $n_g = 1$; their intersections with the plots indicate the principal refractive indices ($n_o$, $n_e$) of o- and e-light in the crystal, as marked by green and pink cross symbols, respectively, while the purple bars denote the superluminal propagation regimes of the ST light.

We plot the birefringent behavior in terms of $n_g$, $n_{og}$, and $n_{eg}$ based on **Eq. (2)**, and analyze the ST birefringence at normal incidence as light enters a YVO₄ crystal from free space, with the OA along the $x$-, $y$-, and $z$-axis (**Figure 2D-F**). An intriguing feature of ST birefringence is the existence of a critical spectral tilt angle $\theta_c$ (marked by red circles in **Figure 2D-F**), at which the group indices of $n_{og}$ and $n_{eg}$ become identical: $\theta_c = \text{arccot}(n - n_o^2/n)$ for OA along the $x$-axis (**Figure 2D**); $\theta_c =$

arccot($n$+$n_o n_e$/$n$) for OA along the $y$-axis (**Figure 2E**); and $\theta_c$ = arccot($n$) for OA along the $z$-axis (**Figure 2E**), where $n$ is the refraction index in free space. At $\theta_c$, the dispersion relations for both polarization components are balanced, allowing the two orthogonally polarized modes to co-propagate without relative delay. This enables synchronized evolution of the o- and e-ST components, beneficial for applications requiring coherent control or precise temporal overlap of different polarization states. It's also observed that when the OA is along the $x$-axis, $\theta_c$ falls within the subluminal ($|n_g|$ > 1)-to-superluminal ($|n_g|$ < 1) transition region (**Figure 2D**); for the $y$-axis alignment, $\theta_c$ resides in subluminal-to-subluminal region (**Figure 2E**); as the OA lies along the $z$-axis, $\theta_c$ locates at the boundary between luminal ($|n_g|$ = 1) and subluminal ($|n_g|$ > 1) regions (**Figure 2F**).

Moreover, when $\theta$ is set below the critical value $\theta_c$, o-ST light propagates faster than its e-ST counterpart; conversely, increasing $\theta$ beyond $\theta_c$ reverses the situation, enabling e-ST light to outpace the o-ST light (**Figure 2D).** In contrast, **Figure 2E, F** show the opposite trend, where o-ST light propagates slower than e-ST light when $\theta$ is below $\theta_c$. Thus, tuning $\theta$ allows e-ST light to propagate faster than o-ST light even in positive crystals along certain crystallographic directions. In negative crystals, the same tuning can slow e-ST light relative to o-ST light. This bidirectional control over relative pulse velocity enables precise temporal ordering of optical signals and markedly differs from conventional light transmission characteristics, where o-light travels always faster than e-light in positive uniaxial crystals (*e.g.*, quartz, MgF$_2$), and vice versa in negative uniaxial crystals (*e.g.*, calcite, LiNbO$_3$).

## 2.3. Experimental validation of space-time birefringence in optical crystals

**Figure 3A** shows the experimental setup for characterization of ST birefringence in the anisotropic medium, *i.e.*, ST light was generated by frequency-doubling the output of a 5-W fs fiber laser with a central wavelength of 1030 nm through second-harmonic generation (SHG). The resulting SHG signal was dispersed using a diffractive grating with a density of 1200 lines/mm, and the spectral phase was controlled via an SLM. The SLM-modulated spectral bandwidth ($\Delta\lambda$) was set to 0.8 nm, while unmodulated spectral components were eliminated through spatial filtering. To characterize the ST birefringence, the initial pulse was split into two optical paths: one served as a reference, the other was used to synthesize ST light, and a custom-cut

YVO$_4$ crystal (10×10×10 mm³, A-cut; $n_o$ = 2.22, $n_e$ = 2.54 measured at $\lambda$ = 515 nm) was introduced into the ST light path (see **Materials and Methods**). The temporal delay between the ST light and the reference pulse was precisely adjusted using a mechanical delay line. Spatially resolved interference patterns were recorded using a CMOS camera mounted on an axially translational stage. The intensity profiles in the ($x$, $y$)-plane of the birefringent ST light and its interference pattern with the reference beam are recorded at various positions along the optical path, as indicated by the dashed ellipses. **Figure 3B** shows a representative spatiotemporal profile of the birefringent ST light, revealing a distinct 'double-X-wave' structure. This profile clearly resolves temporally separated o-and e-ST components, which are delayed by ~11 ps. The generation of birefringent ST light is achieved by normally incident ST light onto a YVO$_4$ crystal, with the polarization oriented at 45° relative to the OA.

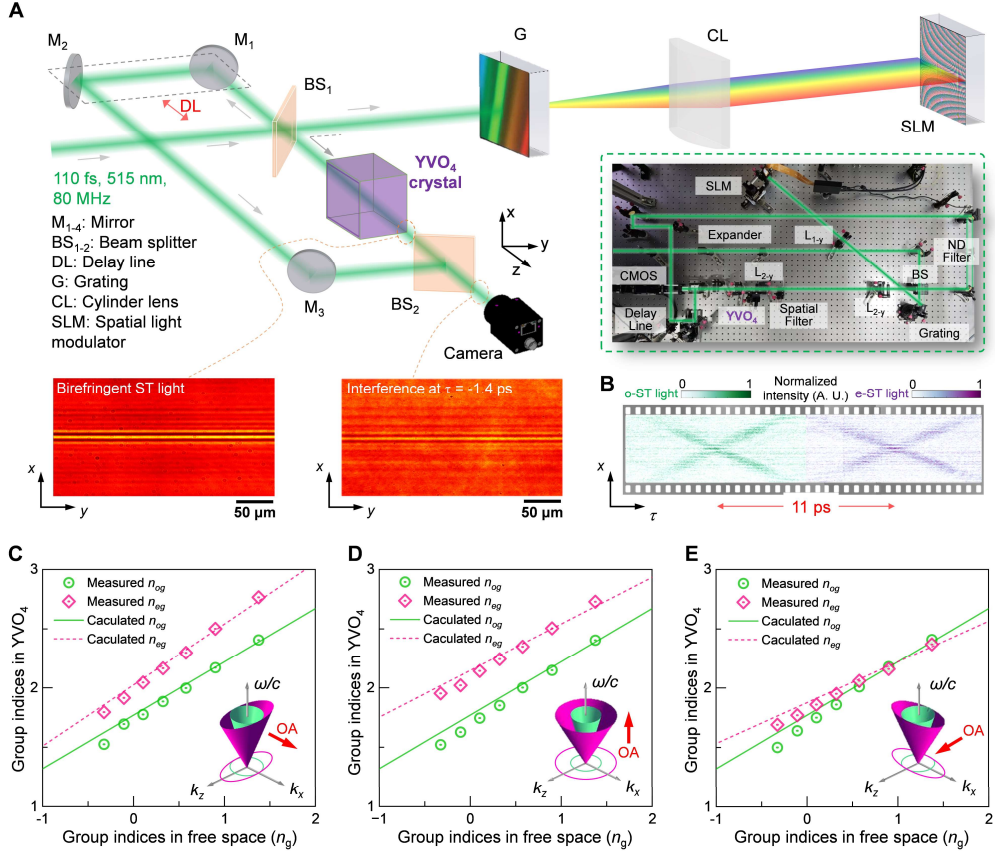

**Figure 3. Synthesis, characterization, and experimental verification of the ST light birefringence in a YVO$_4$ crystal.** (**A**) Schematic of the experimental setup used to synthesize and characterize ST light travelling through a birefringent crystal. A pulsed plane-wave is split into two using a beam splitter (BS$_1$): one path is use to synthesize the ST light by employing a diffraction grating (G), a cylindrical lens (CL), and a spatial light modulator (SLM), while the other serves as the reference beam (**Materials and Methods,** and **Figure S2 and S3**). The inset shows a photograph of the optical setup. The intensity

profiles in $(x, y)$-plane of the birefringent ST light and its interference pattern with the reference beam are recorded at different positions in the optical path, as indicated by the dashed ellipses. (**B**) Representative spatiotemporal profile of a birefringent ST light after passing through a YVO₄ crystal. The time delay between the o-ST (green) and e-ST light (pink) is measured to be ~11 ps. Central wavelength is $\lambda_0 = 515$ nm, $\theta = 36°$, bandwidth is $\Delta\lambda = 0.8$ nm, spectral uncertainty is $\delta\lambda \approx 40$ pm, beam width is 6 μm, and pulse width at the center of the ST light ~1.1 ps. (**C-E**) Measurements of relationship between the free-space group index $n_g$ and the group index of o-ST ($n_{og}$, green) and e-ST ($n_{eg}$, pink) light, with the OA oriented along the $x$-, $y$- and $z$-axis, respectively. Insets in (C-E) show the double-cone representations with OA orientations corresponding to each panel.

We measure the group indices of ST light propagating in YVO₄ crystal under normal incidence for different OA orientations. For each OA orientation, we synthesis ST light with horizontal and vertical polarization separately by adjusting the polarization state of the input fs laser before BS₁, and then measure the group indices of o- and e-ST light independently (**Figures S4, S5**). The corresponding changes in group index upon transitioning from free space to the YVO₄ crystal are presented in **Figure 3C-E**. The results confirm the linear relationship between the group indices in free space ($n_g$) and those within the crystal ($n_{og}$ and $n_{eg}$), as predicted by the birefringence law for ST light illustrated in **Figure 2D-F**, respectively.

## 2.4. Reconfigure the space-time birefringence by phase spectrum programming

Since the group indices of o- and e-ST light in the optical crystal can be tuned by programming the phase spectrum of the incident ST light, this enables precise control over the relative group delay or phase retardation ($\Delta\phi$) between orthogonal polarization states during propagation. We define $\Delta n_{st} = n_{eg} - n_{og}$ as the ST birefringence and derive the corresponding expressions for three special cases:

$$\Delta n_{st} = \begin{cases} n_e - n_o - \frac{n(n_e - n_o)}{n_o^2}(n - \cot\theta), & \text{OA} \, /\!/ \, \vec{x} \\ n_e - n_o + \frac{n(n_e - n_o)}{n_o n_e}(n - \cot\theta), & \text{OA} \, /\!/ \, \vec{y}. \\ \frac{n(n_e^2 - n_o^2)}{n_o n_e^2}(n - \cot\theta), & \text{OA} \, /\!/ \, \vec{z} \end{cases} \quad (3)$$

This formulation reveals controllable birefringence by adjustment of the spectral tilt angle of the incident ST light, without changing the physical properties of the medium or the input wavelength. Given that $\Delta\phi = (2\pi/\lambda)\Delta n_{st}l$ (where $l$ is the crystal thickness), the approach enables reconfigurable phase retardation and dynamic polarization control in birefringent crystals via an all-optical method—eliminating the need for mechanical rotation of optical elements, external electric or magnetic fields, or thermal modulation.

We proceed to validate the above prediction by employing interference measurements to characterize the group delay of the o- and e-ST light with respect to $\theta$. To simultaneously detect the interference fringes for both o- and e-ST components on the CMOS sensor, linearly polarized ST light is incident normally with polarization direction at 45° relative to the OA. The spatiotemporal intensity profile of the birefringent ST light is reconstructed by scanning a 22-ps delay line in the reference arm of the Mach-Zehnder interferometer (**Figure 3A**). The reference pulse delay $\tau$ is first adjusted to achieve spatial and temporal overlap with the birefringent ST light, with high-visibility interference fringes confirming precise alignment. At $\theta = 36°$, the reconstructed spatiotemporal profile of the birefringent ST light shows a clear double-X-wave shape. The pulse widths are measured as 1.1 ps (o-ST) and 1.0 ps (e-ST, full width at half maximum, FWHM), with a corresponding group delay $\Delta\tau = l/v_{eg} - l/v_{og} = 11$ ps. Following this procedure, spatiotemporal profiles are reconstructed and relative group delays are quantified across spectral tilt angles of 36°, 64°, 72°, 84°, and 108°, as summarized in **Figure 4A**. The experimentally measured ST birefringence is thus determined by $\Delta n_{st} = c\Delta\tau/l$. Theoretically, when $\alpha = 90°$ and $\varphi = 45°$, the birefringence expression becomes $n(n - n_g) = n_o(n_o - n_{og}) = \frac{2n_o^2 n_e}{n_o^2 + n_e^2}(n_e - n_{eg})$, yielding $\Delta n_{st} = n_e - n_o - \frac{n(n_e - n_o)^2}{2n_o^2 n_e}(n - \cot\theta)$. We plot both the theoretical and measured ST birefringence values for this case (**Figure S6**), as well as configurations when the OA lies along $x$, $y$, $z$-axis in **Figure 4B-E**, showing good agreements. The tuning range of ST birefringence achieved through this approach is ~0.3, more than 100 times greater than that of conventional crystal birefringence tuning methods, which typically operate within a range of $10^{-5}$ to $10^{-3}$ and rely on external electrical, magnetic, or thermal stimuli[3].

We further analyze the tunable range of $\Delta n_{st}$ and the modulation tunability through adjustments to the spectral phase. When $\alpha = 90°$ and $\varphi = 45°$, the slope $d(\Delta n_{st})/d(\theta)|_{\theta=90°} = (n_o - n_e)/(2n_o^2 n_e) \approx -0.004$ rad$^{-1}$ indicates tuning of birefringence albeit with a limited tuning range, due to constraints imposed by the achievable spectral tilt angle, which is restricted by the SLM resolution. According to **Eq. (3)**, when the OA aligns with the $x$-, $y$-, and $z$-axes, yielding slopes of $d(\Delta n_{st})/d(\theta)|_{\theta=90°} = -n(n_e - n_o)/n_o^2 \approx -0.065$ rad$^{-1}$ (**Figure 4B**), $d(\Delta n_{st})/d(\theta)|_{\theta=90°} = n(n_e - n_o)/(n_o n_e) \approx 0.057$ rad$^{-1}$ (**Figure 4C**), and $d(\Delta n_{st})/d(\theta)|_{\theta=90°} = n(n_e^2 - n_o^2)/(n_o n_e^2) \approx 0.106$ rad$^{-1}$ (**Figure 4D**), respectively. These steeper slopes indicate a broader modulation range, confirming that phase retardation can be effectively reconfigured solely by programming the spectral phase of input ST light.

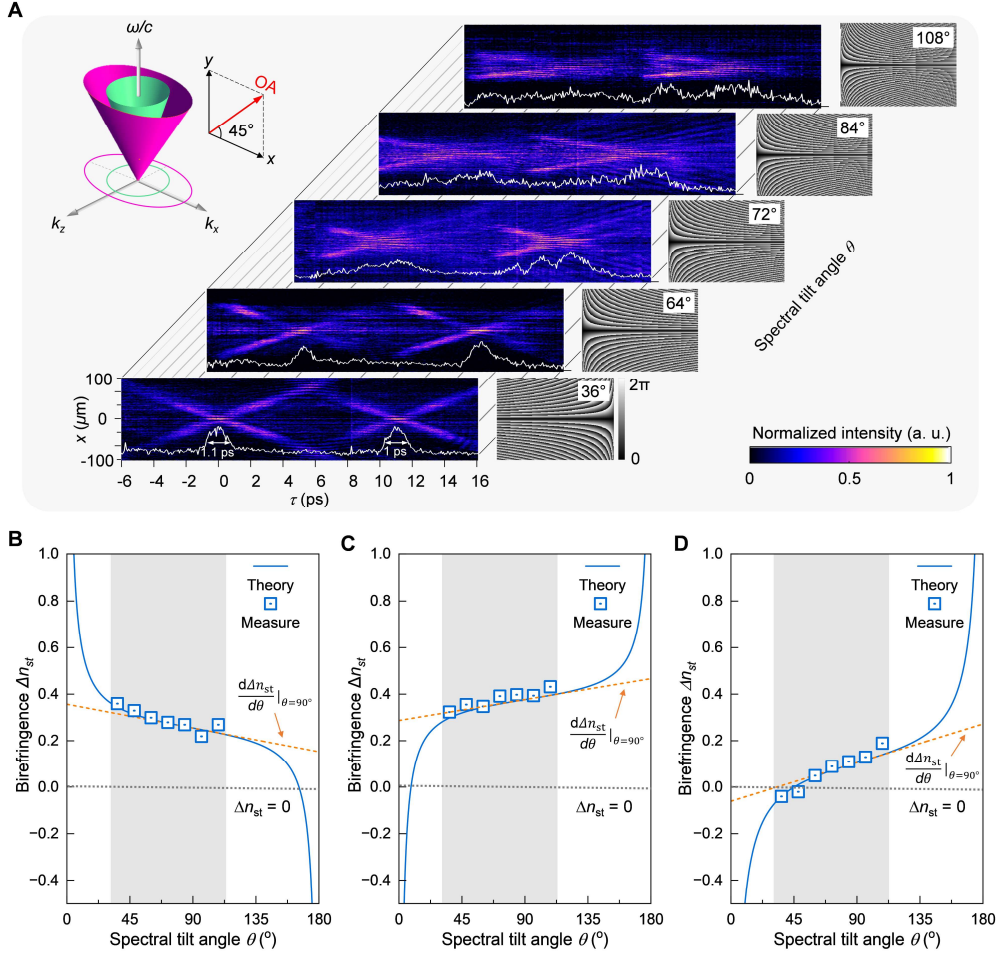

**Figure 4. Experimental verification of reconfigurable ST birefringence in a YVO₄ crystal.** (**A**) Spatiotemporal measurements of o- and e-ST light with different optical path differences after YVO₄ crystal by varying the spectral tilt angle of the incident ST light at 36°, 64°, 72°, 84°, and 108°, respectively. The 2D SLM phase distributions to synthesize ST light with different spectral tilt angles are shown correspondingly. The ST light was normally incident from free space onto a YVO₄ crystal, with the crystal's OA (red arrow) oriented at 45° relative to the $x$-axis in the $(x, y)$-plane (see **Materials and Methods**). The white lines represent the pulse intensity profiles at $x = 0$, $I(0, \tau)$, corresponding to each $\theta$. (**B-D**) Theoretical (blue curves) and experimentally measured (blue square) birefringence values of ST light under varying $\theta$ with the OA oriented along the $x$-axis (B), $y$-axis (C), $z$-axis (D), respectively. The orange dashed lines in each panel indicate the slope of the theoretical curve at $\theta = 90°$, reflecting the modulation tunability of ST birefringence for each OA orientation. The gray dotted lines, corresponding to $\Delta n_{st} = 0$, denote the critical angles at which the birefringence is cancelled for each respective OA orientation.

## 2.5. Space-time birefringence of an obliquely incident ST light

We finally evaluate the ST birefringence under oblique incidence into the YVO₄ crystal. As shown in **Figure 5A-C**, we explicitly illustrate the oblique incidence for the OA oriented along the *x*-, *y*-, and *z*-directions, respectively, where $\phi_1$ is the angle of incidence, $\phi_2$ and $\phi_3$ are the refractive angle of the o- and e-ST light within the crystal, respectively, satisfying the relation: $n\sin\phi_1 = n_o\sin\phi_2 = n_e\sin\phi_3$. The refraction invariant remains valid for o-ST light under oblique incidence and takes the form $n(n-n_g)\cos^2\phi_1 = n_o(n_o-n_{og})\cos^2\phi_2$, whereas the invariant for e-ST light must be modified to account for the anisotropic dispersion effects (see **Text S2** and **Figure S7** in the Supporting Information). Thus, we have the relationship among $n_g$, $n_{og}$ and $n_{eg}$ as given by

$$n(n-n_g)\cos^2\phi_1 = n_o(n_o-n_{og})\cos^2\phi_2 = \begin{cases} \frac{n_e^2(\phi_3)}{n_e(90°-\phi_3)}[n_e(90°-\phi_3)-n_{eg}]\cos^2\phi_3, & \text{OA}//\vec{x} \\ n_e(n_e-n_{eg})\cos^2\phi_3, & \text{OA}//\vec{y} \\ \frac{n_e^2(90°-\phi_3)}{n_e(\phi_3)}[n_e(\phi_3)-n_{eg}]\cos^2\phi_3, & \text{OA}//\vec{z} \end{cases} \quad (4)$$

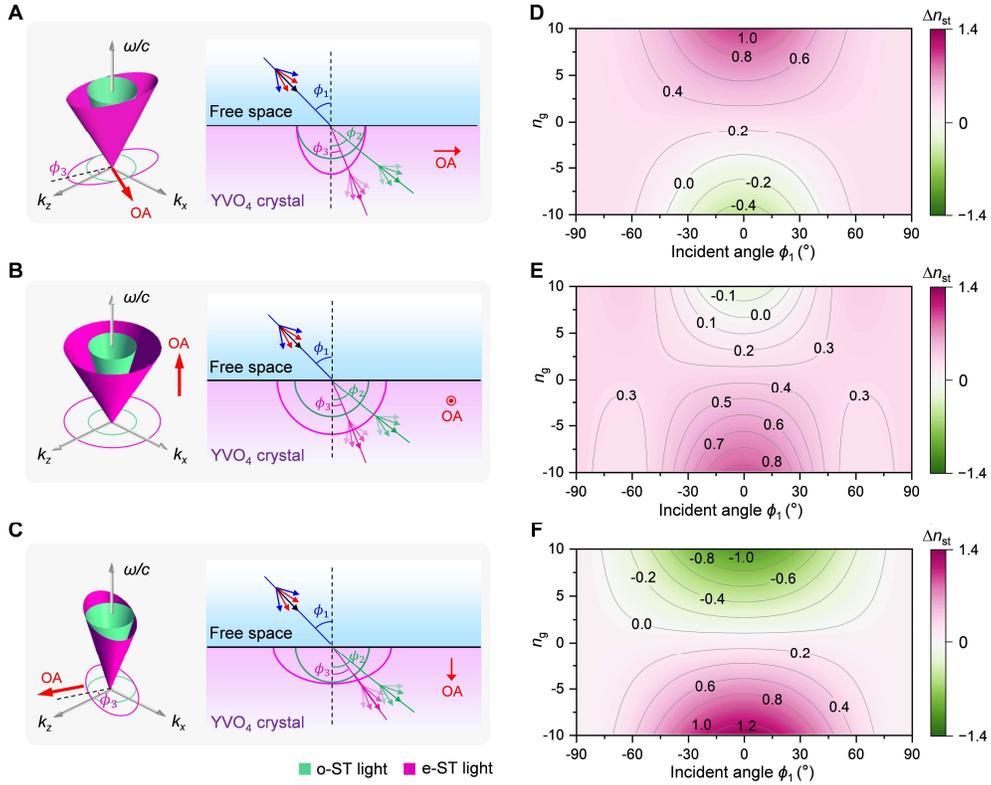

**Figure 5. Birefringence of the ST light obliquely incident from free space to the uniaxial crystal.** (**A**-**C**) Double-light cone representation and schematic illustrations of the ST light obliquely incident at the interface from free space (blue) to a uniaxial positive crystal (purple) with the OA along the *x*- (**A**), *y*- (**B**) and *z*-axis (**C**), respectively. $\phi_1$, $\phi_2$ and $\phi_3$ are the incident angle, refractive angles for the o-ST and e-ST light, respectively. The blue, green and pink arrows indicate the wave vectors of ST light in free space, o-ST light and e-ST light, which are governed by the projection of the double light-cone (ST in free space: blue, o-ST: green, e-ST: pink), respectively. (**D**-**F**) Relationship of the birefringence ($\Delta n_{st}$) with respect to the incident group index ($n_g$) and the incident angle ($\phi_1$), with the crystal's OA orientated along the *x*- (D), *y*- (E) and *z*-axis (F), respectively. The plot is based on **Eq. (4)**. Cancellation of the birefringence of the ST light can occur under specific

incident angle ($\phi_1$) and group index ($n_g$), as indicated by the contour line of $\Delta n_{st} = 0$ on each figure. These contour lines also mark the transition of the ST light between positive and negative birefringence, or vice versa, when obliquely incident into the YVO$_4$ crystal.

The above equations indicate that the group index of o- and e-ST light depends on both the angle of incidence $\phi_1$ and the spectral tilt angle $\theta$ of the incident ST light. Thus, the ST birefringence $\Delta n_{st}$ at oblique incidence can be determined and is plotted in **Figure 5D-F**, accordingly. In all configurations, a zero delay between o- and e-ST light can be achieved by appropriately adjusting $\phi_1$ and $\theta$, as indicated by the contour where $\Delta n_{st} = 0$. This condition corresponds to matched group velocities, thus eliminating temporal walk-off. The tunability allows for synchronization of orthogonally polarized components despite their distinct propagation characteristics, which is particularly advantageous in applications involving ultrafast optics, polarization-insensitive devices, and coherent control. Notably, ST birefringence can be continuously adjusted in a broad range spanning negative to positive values—regardless of the optical crystal's intrinsic optical sign—through precise control of the incidence angle and spectral tilt angle, demonstrating that angular and spectral tuning enable flexible and material-preserving birefringence engineering.

### 3. Discussion and Conclusion

Extending the investigation of ST light to anisotropic media not only deepens our understanding of the light field dynamics in space and time, but also lays the foundation for controlling birefringence through a new degree of freedom. This analysis entails wave manipulation in tensor permittivity environments and challenges the conventional assumption of separable spatial and temporal behavior.[1] In contrast to conventional birefringent elements that rely on fixed material birefringence or mechanical reconfiguration, ST birefringence can be effectively tuned by engineering the internal phase spectrum of input ST light, enabling all-optical, real-time control of polarization dynamics. This approach allows precise manipulation of group velocities without physical movement or alteration of the medium, facilitating accurate pulse timing and inter-wave-packet delays. Moreover, since the mechanism relies on dispersion engineering rather than on absorption or intensity-dependent nonlinearities,[34] it operates efficiently at low optical intensities and over broad bandwidths, highlighting the potential for dynamic controlling wave propagation in complex media.

Tailoring ST light to navigate and exploit anisotropy could enable new paradigms for ultrafast optical control and information processing. It holds particular promise for the development of advanced optical devices with precisely engineered anisotropic properties for controlling light with unprecedented precision in birefringent media. For instance, space-time tailoring permits on-the-fly manipulation of polarization states, facilitating high-speed optical switching for next-generation communications and quantum information protocols.[35] It also benefits adaptive systems requiring dynamic correction—such as deep-tissue imaging,[36, 37] atmospheric transmission, and laser machining[38, 39]—by aligning structured light with local anisotropy to reduce distortions and improve focusing without relying on feedback or deformable optics. Additionally, it allows synchronized excitation in birefringent media by pre-shaping pulses to compensate for temporal walk-off between polarization components, sustaining nonlinear interactions over long distances, significantly enhancing efficiency in wavelength conversion. This ensures simultaneous arrival at the target plane, enabling coherent control of nonlinear processes like second/high harmonic generation,[40] coherent Raman scattering [41] and optical parametric amplification.[42]

While in this study we concentrate on uniaxial optical crystals to enable a more intuitive demonstration, biaxial crystals possess multiple optical axes and exhibit more intricate conical refraction phenomena between ST light and the crystal lattice, leading to sophisticated polarization-dependent effects and more varied spatiotemporal characteristics.[3, 43] Those analysis of such phenomena requires advanced theoretical models and sophisticated experimental setups that are currently under development and fall outside the scope of the present work. Meanwhile, understanding how ST light interacts with anisotropic media is crucial from both fundamental and applied perspectives, as it fundamentally explores the limits of wave packet control in environments where light-matter interactions are governed by tensor permittivity rather than scalar values, and tests foundational principles of light evolution under strong spatial and temporal non-separability.[30] In such scenarios, the traditional assumptions of separable spatial and temporal dynamics no longer hold, requiring new frameworks to accurately describe and predict light behavior.

Overall, the exploration of ST light propagation in anisotropic materials is a critical endeavor sitting at the intersection of fundamental wave physics and broad cutting-edge implications for optics and photonics. Unlocking their full potential promises to catalyze advancements across optical signal processing,[44] optical neuromorphic computing for artificial intelligence[45, 46] and photonic quantum processing,[35, 47] shaping the future of compact, reconfigurable photonic devices and optical systems.

## 4. Materials and Methods

**Synthesis of ST light.** A vertically polarized pulsed beam with a central wavelength of 515 nm, a pulse duration of approximately 180 fs and a repetition rate of 80 MHz, generated by second-harmonic generation (SHG) from a fiber laser (Fibre Series, Keyun Photoelectric Technology, China), is expanded using a Galilean telescope—comprising a concave lens ($f = -30$ mm) and a convex lens ($f = 150$ mm)—to closely approximate a plane wavefront. The beam's spectrum is spatially dispersed via a diffraction grating (1200 lines/mm, Lbtek BG25-1200-1000, Shenzhen LUBON Technology, China) at an incidence angle of ~17.83°, with the second diffraction order observed at 68.40°, resulting in an angular dispersion of 0.0065°/nm. Following this, the diffracted beam is collimated in a $2f$ configuration by a cylindrical lens ($L_{1-y}$, $f = 400$ mm), ensuring that each wavelength is mapped onto a distinct vertical line (along the $x$ direction) of a polarization-sensitive spatial light modulator (SLM; HDSLM80R Plus, UPOLabs, China). This arrangement enables precise, wavelength-dependent phase modulation $\exp[i\psi(x, y)]$. Upon reflection, the beam retraces its path through $L_{1-y}$ and the grating G, thereby establishing a correlation between temporal and spatial frequencies and forming a structured ST wave packet (see **Figure S2** for details).

**Phase modulation of SLM.** As shown in **Figure S2**, each wavelength $\lambda$ is mapped to a corresponding position $y(\lambda)$ on the SLM through the diffraction grating G, where a 2D phase modulation $\exp[i\psi(x, y)]$ is applied. To establish the desired $k_x$–$\lambda$ correlation, a linear phase gradient $\psi(x, y) = (|x-x_c|/x_s)(y/\alpha y_s)^{1/2}$ is introduced along the $x$-direction at each $y(\lambda)$, with $\alpha$ defines the curvature of the hyperbolic ST spectrum, $x_s$ and $y_s$ are scaling factors depending on the SLM details (pixel size and wavelength spread across the SLM), and $x_c$ is the central point on the SLM along the vertical direction. Since each wavelength in ST light possesses two symmetric spatial frequencies, $+k_x$ and $-k_x$, the SLM is divided into two distinct regions: one encodes the positive spatial frequency ($+k_x$), and the other encodes the negative counterpart ($-k_x$). To maximize the utilization of SLM pixels, the initial spatial frequencies are reduced by a factor of four (to $\pm k_x/4$) and subsequently restored via angular magnification using two relay lenses ($L_{2\text{-}x}$ and $L_{3\text{-}x}$).

***Characterization of the time-averaged intensity profile.*** The beam reflected by the SLM retraces its incident path and, after passing through a grating, is directed by a beam splitter ($BS_1$) into a $4f$ system composed of lenses $L_{2\text{-}x}$ ($f = 400$ mm) and $L_{3\text{-}x}$ ($f = 100$ mm). This $4f$ system images the ST light with a $4\times$ demagnification along the $x$-direction. Due to the limited diffraction efficiency (~95%) of the SLM, unwanted zero-order components can interfere with the observation of the ST light and are therefore eliminated using a spatial filter (SF) placed in the Fourier plane between $L_{2\text{-}x}$ and $L_{3\text{-}x}$, where unmodulated plane waves propagating along the $z$-direction are spatially separated from the ST light. Finally, a scanning CMOS camera (STC-MBA1002POE, Shenzhen LUBON Technology, China) captures the beam profile along the $z$-axis $I(x,z) \propto \int \mathrm{d}t |E(x,z;t)|^2$, revealing the diffraction-free propagation length of the ST light.

***Characterization of the spatiotemporal spectrum of ST light***. The spatiotemporal spectrum of the ST light is obtained by applying the Fourier transform in both the temporal and spatial domains $\tilde{E}(k_x, \omega) = \iint \mathrm{d}x\mathrm{d}t E(x,0,t)e^{-i(k_x x - \omega t)}$. A second beam splitter ($BS_2$) directs the synthesized ST light into a separate path prior to the second grating pass, enabling time-domain Fourier transformation. A spherical lens ($L_{1\text{-}s}$, $f = 75$ mm) subsequently performs the spatial Fourier transformation, with the ST spectrum captured at the focal plane of the lens with a CMOS camera. The direct current (DC) components are preserved and clearly visible, as shown in **Figure S2**.

***Characterization of the group index of ST light in free space***. As shown in **Figure S3**, to measure the group index of ST light, a reference pulse is split from the original beam using beam splitter $BS_1$, and its power is adjusted via a neutral density filter. This reference pulse is then made to interfere with the ST light after propagating through an optical delay line. Maximum fringe visibility is observed on the CMOS camera when the two pulses are temporally and spatially overlapped. By translating the camera over a distance $L$ in free space, path differences arise due to the differing group indices—$n_g$ for ST light and $n = 1$ for the reference pulse—resulting in reduced interference visibility, the optical path difference (OPD) is $(n_g - n)L$. To compensate for this temporal mismatch, the optical delay line is adjusted by a displacement $\Delta l$, thereby restoring high-visibility interference fringes. This procedure enables spatially resolved characterization of the pulse overlap. The group velocity and corresponding group index of ST light in free space are subsequently determined based on the measured values of $L$ and $\Delta l$, $(n_g - n)L = \Delta l$, so that $n_g = n + \Delta l/L$.

***Characterization of the group indices of the birefringent ST light.*** The o-ST and e-ST components are measured independently. The polarization of the ST light is controlled using a SLM in conjunction with other optical elements prior to the grating to change the polarization of the original laser. Introducing a custom YVO₄ crystal of length $L$ into the ST light path induces a group delay, which reduces fringe visibility by the optical path difference between the reference and ST light, OPD = $(n_{e,og} - n_g)L$. By adjusting the delay line by $\Delta l$, temporal overlap is restored, recovering high-visibility interference fringes and enabling accurate determination of the crystal's group velocity and refractive index, $(n_{e,og} - n_g)L = n\Delta l$, $n_{e,og} = n_g + n\Delta l/L$.

***Characterization of spatiotemporal intensity profile of the birefringent ST light***. The visibility of spatially resolved interference fringes is proportional to $|\psi(x, \tau)|^2$, where $\psi(x, \tau)$ is the envelope function of ST light, enabling reconstruction of the spatiotemporal intensity profile through interferometric measurements. As the optical delay is scanned in the area where the ST light and reference beam overlap, the fringe visibility varies correspondingly. To precisely capture this variation, the delay line is incremented in 0.01 mm steps, while a high-resolution nano-positioning stage (WLS-50, Nators, China) applies fine adjustments of 75 nm at each step to enhance temporal resolution. At every delay position, a CMOS camera acquires 15 interferograms to ensure a robust determination of the fringe visibility $v = (I_{max} - I_{min})/(I_{max} + I_{min})$, with each dataset corresponding to a single spatial line in the reconstructed intensity profile (**Figure S5**). To simultaneously measure the intensity profiles of the o-ST and e-ST components, a YVO₄ crystal is oriented at 45° relative to the incident polarization. Upon transmission through the crystal, temporal dispersion separates the o-ST and e-ST components, resulting in a distinct 'double X-wave' structure.

## Acknowledgements

We thank Drs. Cheng Wang and Xiaoming Lu for their assistance with the maintenance of the fs laser system, Dr. Zongxin Zhang for valuable discussions on the spatiotemporal characterization of the fs laser pulses, and Dr. Chengchun Zhao for providing the optical crystals and for insightful discussions on crystal optics.

## Data Availability Statement

All data are available in the main text or the Supporting Information. Additional data related to this paper may be requested from the authors.

## Conflict of Interest

The authors declare no conflicts of interest.